\documentclass[12pt,preprint]{aastex}
\usepackage{amsmath}

\shorttitle{Disk rotation curves for arbitrary density profiles}
\shortauthors{Pierens \& Hur\'e}

\begin{document}

\title{Rotation curves of galactic disks for arbitrary surface density profiles: a simple and efficient recipe}

\author{Arnaud Pierens}
\affil{LUTh/Observatoire de Paris-Meudon (UMR 8102 CNRS),\\ Place Jules
Janssen, 92195 Meudon Cedex, France}

\email{arnaud.pierens@obspm.fr}

\and

\author{Jean-Marc Hur\'e}
\affil{LUTh/Observatoire de Paris-Meudon (UMR 8102 CNRS),\\ Place Jules
Janssen, 92195 Meudon Cedex, France}
\affil{Universit\'e Paris 7 Denis Diderot,\\ 2 Place Jussieu, 75251 Paris Cedex 05, France}
\email{jean-marc.hure@obspm.fr}

              
\begin{abstract}
There is apparently a widespread belief that the gravitational field (and subsequently the rotation curve) ``inside'' razor-thin, axially symmetric disks can not be determined accurately from elliptic integrals because of the singular kernel in the Poisson integral. Here, we report a simple and powerful method to achieve this task numerically using the technique of ``density splitting''.

\end{abstract} 


\section{Introduction. Statement of the problem}
One of the oldest problem in galactic dynamics is the determination of
the mass distribution in the stellar and gaseous disk from
observed velocities. Observationally, noisy measurements along with a certain velocity dispersion makes this inverse problem difficult to solve (Beauvais \& Bothum, 2001). From a theoretical point of view, various levels of approximation are possible. For instance, the inversion is straightforward if the disk is considered as a flattened spheroid, since the surface density is simply obtained by an inverse Abel transform over the velocity field (Brandt, 1960). Flattened spheroids are however a poor representation of real disks (e.g. Kochanek 2001). Another, more intuitive way to proceed is to generate a collection of surface density profiles, then to determine the corresponding potential or gravitational field, and find only those which model the velocity data at best. For a razor-thin axi-symmetrical disk, only one numerical quadrature over its radial extent must be performed to obtain the potential or the field, which is a priori very convenient because time in-expansive. Meanwhile, this operation is commonly believed to be numerically uncomfortable when 
 using elliptic integrals (Binney \& Tremaine, 1987, Cuddeford, 1993, Kochanek, 2001). The reason is that these functions contain a singularity everywhere in the source. As suggested by Binney \& Tremaine (1987), this drawback can be simply avoided by considering field points located just above/below the equatorial plane. Even if the potential and its radial derivative do generally not exhibit strong variations slightly off the mid-plane, such a vertical shift introduces an inconsistency or a bias. As a matter of fact, instead of elliptic integrals, most authors work with Bessel functions which have finite amplitudes (Toomre, 1963). But their oscillatory behavior and specially their spatial extension (the infinite range) are a severe disadvantage from a numerical point of view (Cuddeford, 1993; Cohl \& Tohline 1999).\\

In this paper, we show that point mass singularities can be properly and easily handled numerically from elliptic integrals by density splitting. For the case of a disk with zero thickness as considered here, the gravitational potential and field can be determined exactly in the equatorial plane without any shift, and whatever the surface density profile (provided it vanishes at the inner and outer edges of the disk). The extension to tri-dimensional disks is possible.

\section{The surface density splitting method}

For a disk with zero thickness, the expression for the radial gravitational field in the equatorial plane is known in a closed form. Using cylindrical coordinates, this reads (e.g Durand, 1964):
\begin{equation}
g_r(r)=\frac{G}{r}\int_{r_{\rm in}}^{r_{\rm out}}\sqrt\frac{a}{r}k \Sigma(a)\left[\frac{{\bf E}(k)}{\varpi}-{\bf K}(k)\right]da,
\label{eq:potential}
\end{equation}

where $\Sigma$ is the total surface density, $r$ is the field point, $a$ refers to the material distribution, $r_{\rm in}$ and $r_{\rm out}$ are respectively the inner and the outer radius of the disk, and $k$ and $\varpi$ are respectively defined  by
\begin{equation}
k=\frac{2\sqrt{ar}}{(a+r)}
\end{equation}

and
\begin{equation}
\varpi=\frac{a-r}{a+r}
\end{equation}

Finally, ${\bf K}$ (respectively ${\bf E}$) is the complete elliptic integral of the first (resp. second) kind. Note that the integrand of equation (\ref{eq:potential}) diverges as the modulus $k\rightarrow 1$, namely when $a\rightarrow r$. The nature of the singularity is twofold: a logarithmic singularity due to the ${\bf K}$-function, and a hyperbolic one due to the term $1/\varpi$. However, this singularity is integrable. There are different ways to estimate improper integrals. Here is a simple and efficient recipe. Let us write the surface density as
\begin{equation}
\Sigma(a)=\Sigma(r)+\delta \Sigma(a,r),
\end{equation}

where $\Sigma(r)$ is the surface density at the field point and $\delta \Sigma(a,r)$ is the ``residual'' surface density which, by construction, equals zero when $a=r$. Thus, equation (\ref{eq:potential}) can be written as the sum of two
contributions, namely:
\begin{equation}
g_r(r)=g_r^{\rm homo.}(r)+g_r^{\rm res.}(r),
\end{equation}

where
\begin{equation}
g_r^{\rm homo.}(r)=\frac{G\Sigma(r)}{r}\int_{r_{\rm in}}^{r_{\rm out}}\sqrt\frac{a}{r}k\left[\frac{{\bf E}(k)}{\varpi}-{\bf K}(k)\right]da
\label{eq:hom}
\end{equation}

is the radial gravitational field due to a radially homogeneous
disk and
\begin{equation}
g_r^{\rm res.}(r)=\frac{G}{r}\int_{r_{\rm in}}^{r_{\rm out}}\sqrt\frac{a}{r}k \;
\delta \Sigma(a,r)\left[\frac{{\bf E}(k)}{\varpi}-{\bf K}(k)\right]da
\label{eq:1}
\end{equation}

corresponds to the residual profile. The point is that the expression for $g_r^{\rm homo.}$ also exists in a closed analytical form (see Appendix \ref{sec:appenA}). This quantity is finite everywhere inside the disk, even at the disc edges provided that $\Sigma(r)$ vanishes continuously there. It is important to mention that this restriction is no more necessary if we consider the disk as a tri-dimensional system. Besides, $g_r^{\rm res.}$  can be easily computed numerically because both $\delta\Sigma(a,r){\bf K}(k)$ and $\delta\Sigma(a,r){\bf E}(k)/\varpi$ are fully regular whatever $a$ and $r$, and especially for $a=r$ (see Appendix  \ref{sec:appenB} for a proof). As a consequence, the accuracy on  $g_r^{\rm res.}$ (and subsequently on the total field $g(r)$) depends only on the performance of the quadrature scheme used to performed the radial integral in equation (\ref{eq:1}).

\begin{figure}
\epsscale{0.9}
\plotone{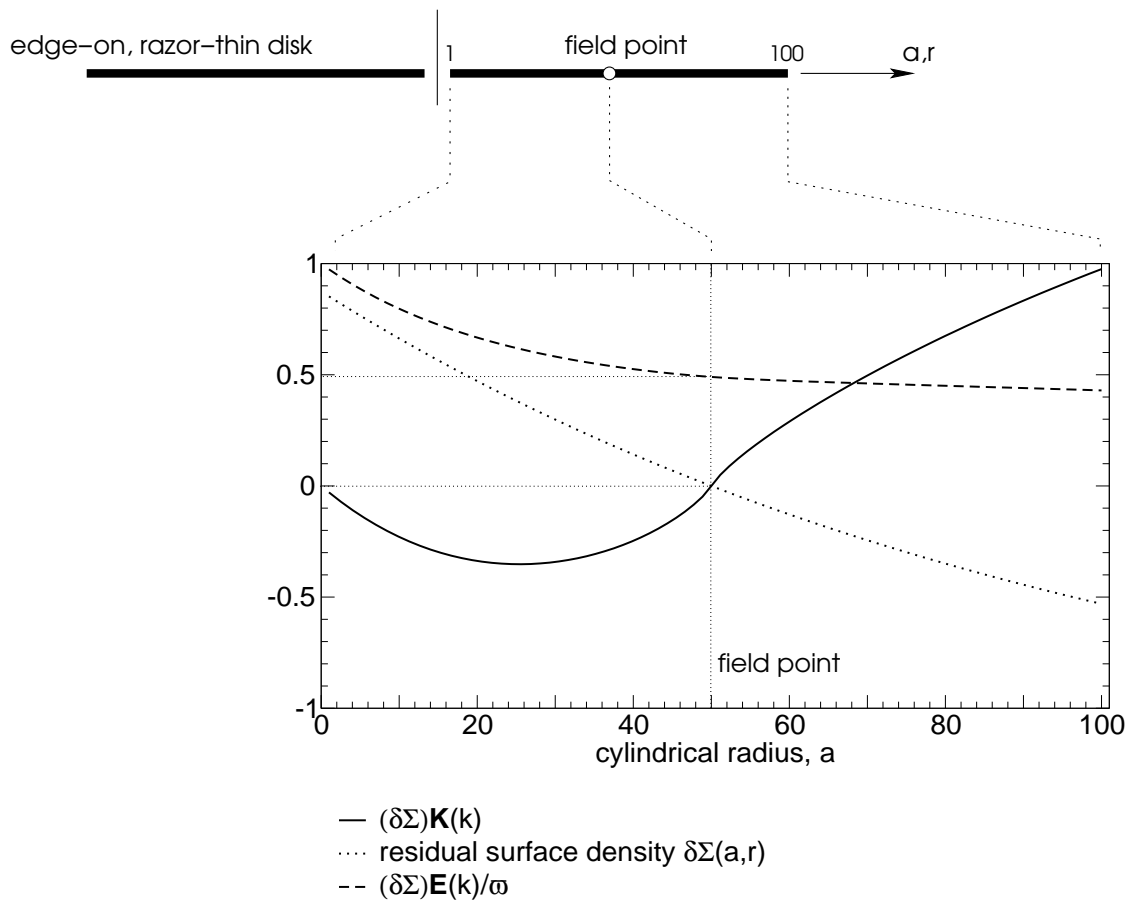}
\caption{Residual surface density profile $\delta\Sigma(a,r)$, and functions $\delta\Sigma(a,r){\bf K}(k)$ and $\delta\Sigma(a,r){\bf E}(k)/\varpi$ appearing in Eq.(\ref{eq:1}) (re-scaled for clarity), for a razor-thin disk extending from $r_{\rm in}=1$ to $r_{\rm out}=100$, and with surface density profile $\Sigma\propto \exp(-r/r_{\rm out})$. The field point is located at $r=50$.}
\label{figure}
\end{figure}

In figure \ref{figure}, we illustrate this simple and efficient technique by considering a disk with inner edge $r_{\rm in}=1$, outer edge $r_{\rm out}=100$ and surface density $\Sigma(r)\propto \exp(-r/r_{\rm out})$ as commonly used to model spiral galaxies (Freeman, 1970). The graph shows the residual profile $\delta\Sigma(a,r)$, as well as the two regular functions $\delta\Sigma(a,r){\bf K}(k)$ and $\delta\Sigma(a,r){\bf E}(k)/\varpi$. Here, the field point where stands the singularity is arbitrarily set to $r=50$ (about the middle of the disk).

\section{Rotation curve of the disk}

We can now apply this method to compute the rotation law $\Omega(r)$ of any disk, whatever its size and surface density profile. Here, we simply assume that the contribution to the rotation is only gravitation (and neglect pressure and viscosity effects\footnote{According to Sofue \& Rubin (2001), this is a good approximation because of the small velocity dispersion of interstellar gas.}). Thus, $\Omega(r)$ is defined by the relation $\Omega^2(r)=\Omega_{\rm homo.}^2(r)+\Omega_{\rm res.}^2(r)$ where:
\begin{equation}
\Omega_{\rm homo.}^2(r)=-\frac{g_r^{\rm homo.}}{r}
\end{equation}

and
\begin{equation}
\Omega_{\rm res.}^2(r)=-\frac{g_r^{\rm res.}}{r}
\end{equation}
Using the previous expressions  for the radial field, these two terms respectively read:
\begin{equation}
\Omega_{\rm homo.}^2(r)=-\frac{4G\Sigma(r)}{r}\left\{\frac{r_{\rm out}}{r}\left[{\bf E}\left(\frac{r}{r_{\rm out}}\right)-{\bf K}\left(\frac{r}{r_{\rm out}}\right)\right]-{\bf E}\left(\frac{r_{\rm in}}{r}\right)+{\bf K}\left(\frac{r_{\rm in}}{r}\right)\right\}
\label{eq:omegahomo}
\end{equation}

and
\begin{equation}
\Omega_{\rm res.}^2(r)=-\frac{G}{r^2}\int_{r_{\rm in}}^{r_{\rm out}}\sqrt\frac{a}{r}k\delta \Sigma(a,r)\left[\frac{{\bf E}(k)}{\varpi}-{\bf K}(k)\right]da
\label{eq:2}
\end{equation}

These two expressions are easy to compute numerically, for reasons mentioned above. Let us remind that at the edges, $\Omega_{\rm res.}^2(r)$ vanishes because $\Sigma(r)=0$ as assumed (see above).

\section{Concluding remarks}

We have reported a simple and efficient method to compute the radial field and the rotation curve of any razor-thin axi-symmetric disk in the mid-plane, using complete elliptic integrals.
As a matter of fact, a similar treatment can be considered for the gravitational potential. Even, this splitting technique can be employed to disks with finite (non-zero) thickness as suggested in Hur\'e (2003), and to the fully tri-dimensional systems as well (Pierens \& Hur\'e 2004).

\acknowledgments

\appendix

\section{Radial field due to a a razor-thin disk with constant surface density}
\label{sec:appenA}
Let us begin with the expression for the radial field due to a disk with constant surface density, i.e. equation (\ref{eq:hom}).
This expression can be written as follows:
\begin{equation}
g_r^{\rm homo.}(r)=\frac{G\Sigma(r)}{r}\left\{\int_{r_{\rm in}}^{r}\sqrt\frac{a}{r}k\left[\frac{{\bf E}(k)}{\varpi}-{\bf K}(k)\right]da+\int_{r}^{r_{\rm out}}\sqrt\frac{a}{r}k\left[\frac{{\bf E}(k)}{\varpi}-{\bf K}(k)\right]da\right\}
\label{eq:grhom}
\end{equation}

We can now set $u=\frac{a}{r}$ in the first integral and $v=\frac{r}{a}$ in
the second one. Changing the modulus in the elliptic integrals according to (Gradshteyn \& Ryzhik, 1980):
\begin{equation}
{\bf K}\left(\frac{2\sqrt{u}}{1+u}\right)=(1+u)\;{\bf K}(u),
\end{equation}

and
\begin{equation}
{\bf E}\left(\frac{2\sqrt{u}}{1+u}\right)=\frac{1}{1+u}\left[2{\bf E}(u)-{u^\prime}^2{\bf K}(u)\right]
\end{equation}

with ${u^\prime}^2=1-u^2$ and $u < 1$, then equation (\ref{eq:grhom}) becomes:
\begin{equation}
g_r^{\rm homo.}(r)=-4G\Sigma\left[\int_{r_{\rm in}/r}^{1}\frac{u{\bf E}(u)}{{u^\prime}^2}du+\int_{1}^{r/r_{\rm out}}\frac{{\bf E}(v)-{v^\prime}^2{\bf K}(v)}{v^2{v^\prime}^2}dv\right]
\label{eq:grhom2}
\end{equation}

Each integral in this expression is known. Actually, we have (Gradshteyn \& Ryzhik, 1980):
\begin{equation}
\int\frac{u{\bf E}(u)}{{u^\prime}^2}du={\bf K}(u)-{\bf E}(u)
\end{equation}

and
\begin{equation}
\int\frac{{\bf E}(v)-{v^\prime}^2{\bf K}(v)}{v^2{v^\prime}^2}dv=\frac{{\bf K}(v)-{\bf E}(v)}{v}
\end{equation}

Inserting these relations into equation (\ref{eq:grhom2}) and re-arranging terms yields the formula
\begin{equation}
g_r^{\rm homo.}=4G\Sigma\left\{\frac{r_{\rm out}}{r}\left[{\bf E}\left(\frac{r}{r_{\rm out}}\right)-{\bf K}\left(\frac{r}{r_{\rm out}}\right)\right]-{\bf E}\left(\frac{r_{\rm in}}{r}\right)+{\bf K}\left(\frac{r_{\rm in}}{r}\right)\right\}
\end{equation}

\section{Singularity removal}
\label{sec:appenB}
In order to compute numerically equations (\ref{eq:1}) and (\ref{eq:2}), we have to check that $\delta\Sigma(a,r){\bf K}(k)$ and $\delta\Sigma(a,r){\bf E}(k)/\varpi$ remain finite when $a\rightarrow r$, which is not obvious at first glance. Since $\delta \Sigma(r,r)=0$ by construction, a Taylor expansion of $\delta\Sigma$ around $a=r$ yields:
\begin{equation}
\begin{split}
\delta\Sigma(a,r) &=(a-r)\left(\frac{\partial \delta\Sigma}{\partial
  a}\right)_{a=r}+{\cal O}\left((a-r)^2\right) \\
                  &=(a-r)\left(\frac{\partial \Sigma}{\partial
  a}\right)_{a=r}+{\cal O}\left((a-r)^2\right) \\
                  &=(a+r)\varpi\left(\frac{\partial \Sigma}{\partial
  a}\right)_{a=r}+{\cal O}\left((a-r)^2\right)
\label{eq:taylor}
\end{split}
\end{equation}

where $\left(\frac{\partial\Sigma}{\partial a}\right)_{a=r}$ is in general finite everywhere. Since ${\bf E}(k)$ is bounded, this expansion shows
that $\delta\Sigma(a,r){\bf E}(k)/\varpi$ is finite at the field point  (see also figure \ref{figure}), namely
\begin{equation}
\delta\Sigma(r,r)\frac{{\bf E}(1)}{\varpi} = (a+r)\left(\frac{\partial \Sigma}{\partial a}\right)_{a=r}
\end{equation}

Besides, since $ {\bf K}(k) \rightarrow \frac{1}{2}\ln(16/{\varpi}^2)$ as $k\rightarrow 1$ (Abramowitz \& Stegun, 1970), we have $\delta\Sigma(r,r){\bf K}(1) = 0 $ for $a=r$ (see figure \ref{figure} again).

\end{document}